\documentclass[onecolumn,floatfix,superscriptaddress,a4paper,
               showpacs,showkeys,nofootinbib,preprint]{revtex4}
\usepackage{epsfig}
\usepackage{latexsym}
\usepackage{xspace}
\usepackage{hyperref}
\usepackage[latin2]{inputenc}
\usepackage{indentfirst}
\usepackage{enumerate}
\usepackage{color}

\usepackage{amsmath}
\usepackage{amssymb}
\usepackage[english]{babel}
\usepackage{url}

\newcommand*{\dis}{\displaystyle}

\newcommand{\mean}[1]{\langle #1 \rangle}

\newcommand{\Eq}[1]{Eq.~(\ref{#1})}

\newcommand{\der}[2]{\frac{\partial #1}{\partial #2}}
\newcommand\ddfrac[2]{\frac{\displaystyle #1}{\displaystyle #2}}

\newcommand\id[0]{^{\rm id}}

\newcommand\tab[2]{
\begin{center}
\begin{table}
\begin{tabular}#1
\end{tabular}
\caption{#2}
\end{table}
\end{center}}

\newcommand\wt[1]{\widetilde{#1}}

\newcommand{\eq}[1]{\begin{align} #1 \end{align}}

\begin{document}

\title{Quantum van der Waals and Walecka models of nuclear matter}

\author{R. V. Poberezhnyuk}
\affiliation{Bogolyubov Institute for Theoretical Physics, 03680 Kiev, Ukraine}
\author{V. Vovchenko}
\affiliation{Institut f\"ur Theoretische Physik,
Goethe Universit\"at Frankfurt, D-60438 Frankfurt am Main, Germany}
\affiliation{Frankfurt Institute for Advanced Studies, Giersch Science Center,
D-60438 Frankfurt am Main, Germany}
\affiliation{Department of Physics, Taras Shevchenko National University of Kiev, 03022 Kiev, Ukraine}
\author{D. V. Anchishkin}
\affiliation{Bogolyubov Institute for Theoretical Physics, 03680 Kiev, Ukraine}
\affiliation{Frankfurt Institute for Advanced Studies, Giersch Science Center, D-60438 Frankfurt am Main, Germany}
\affiliation{Department of Physics, Taras Shevchenko National University of Kiev, 03022 Kiev, Ukraine}
\author{M. I. Gorenstein}
\affiliation{Bogolyubov Institute for Theoretical Physics, 03680 Kiev, Ukraine}
\affiliation{Frankfurt Institute for Advanced Studies, Giersch Science Center, D-60438 Frankfurt am Main, Germany}

\date{\today}

\begin{abstract}
A comparable study of the quantum van der Waals and Walecka models of  nuclear matter is presented.
Each model contains two parameters which characterize the repulsive and attractive interactions between nucleons.
These parameters are fixed in order to reproduce the known properties of the nuclear ground state.
Both models 
predict a first-order liquid-gas phase transition and a very similar behavior in the vicinity of the critical point.
Critical exponents of the quantum van der Waals model are studied both analytically and numerically.
There are
important differences in the behavior of the thermodynamical functions of the considered models at large values of the nucleon number density.
At the same time both models fall into the universality class of mean-field theory.

\end{abstract}



\maketitle

\section{Introduction}

Nuclear matter is a hypothetical infinite system of interacting nucleons, which has a long history of research.
The known phenomenology of a nucleon-nucleon potential suggests repulsive interactions at a short range and attractive interactions at an intermediate range.
Similarly to most molecular systems, a liquid-gas phase transition takes place in nuclear matter.
Experimentally, the presence of a liquid-gas phase transition in nuclear matter was first reported
in Refs.~\cite{Finn:1982tc,Minich:1982tb,Hirsch:1984yj} by indirect observations, while the direct measurements of the nuclear caloric
curve were first done by ALADIN collaboration~\cite{Pochodzalla:1995xy}, and were
later followed by other experiments~\cite{Natowitz:2002nw,Karnaukhov:2003vp}.
Thermodynamics of nuclear matter and its applications to the production of nuclear fragments in heavy ion collisions were considered in Refs. \cite{Jennings:1982wi,Ropke:1982ino,Fai:1982zk,Biro:1981es,Stoecker:1981za,Csernai:1984hf}
in 1980s (see Ref.~\cite{Csernai:1986qf} for a review of these early developments).

The properties of nuclear matter are described by many different models, in particular by those which employ the relativistic mean-field theory~(RMF)~\cite{Serot:1984ey,Zimanyi:1990np,Brockmann:1990cn,Mueller:1996pm,Bender:2003jk},
as well as parity doubling~\cite{Detar:1988kn,Zschiesche:2006zj,Steinheimer:2011ea}.
In the RMF models interactions are treated in the mean-field approach.
The repulsive interactions are normally mediated through the vector meson exchange, while the attractive interactions are described by the scalar meson exchange.

The classical van der Waals (vdW) equation was recently generalized to include the effects of quantum statistics,
and then applied for a description of the symmetric nuclear matter in Refs.~\cite{Vovchenko:2015vxa,Redlich:2016dpb}.
In the resulting quantum vdW (QvdW) model the repulsive interactions are modeled by means of
the excluded-volume correction, while the attractive interactions are described by the density-proportional mean field.
The QvdW model is fairly good in describing the basic properties of nuclear matter.
A more general vdW-type formalism, which is based on the real gas models of equation of state and which
allows variations on the excluded-volume mechanism and on the attractive mean field, was recently presented in Ref.~\cite{Vovchenko:2017cbu}.
The vdW-type approach is quite different from the conventional nuclear matter
models which are mostly based on the RMF theory, such as the Walecka model~\cite{Walecka:1974qa,Serot:1984ey} and its various generalizations~(see, e.g., Refs.~\cite{Dutra:2012mb,Dutra:2014qga,Li:2008gp} for an overview).
One notable difference is the absence of the effective mass concept in vdW model.
In contrast, the effective mass is one of the main quantities in the RMF approach.
Nevertheless, qualitative features of the nuclear matter phase diagram look very similar in both approaches.

In this work, a comparable study of the Walecka and QvdW models of nuclear matter is presented.
Both models contain two parameters, which characterize repulsive and attractive interactions within the mean field approach.
They are fixed in order to reproduce known properties of a nuclear ground state.
The predictions of the models regarding the location of the critical point and nuclear incompressibility are compared to each other, as well as to the available empirical data.
A {\it hybrid model}, where the attractive interactions are mediated through the scalar meson exchange and where the repulsive interactions are modeled by means of the excluded-volume correction, is considered as well for completeness.

The critical behavior in the vicinity of the critical point (CP) is studied in some detail in the present work.
Critical exponents, which characterize this critical behavior, are studied both analytically and numerically in the QvdW model, and compared to
the predictions of the mean-field universality class.

The paper is organized as follows.
Section \ref{sec:Models} describes the three models under consideration: the Walecka model, the QvdW model, and the Hybrid model.
Section~\ref{sec:nm} presents the predictions of the models regarding the nuclear matter properties.
Section~\ref{sec:cr} discusses the critical behavior of the two models in the vicinity of the CP of nuclear matter.
Summary in Sec.~\ref{sec:summary} closes the article, and Appendix presents the
thermodynamical functions of the ideal Fermi gas.

\section{Models}
\label{sec:Models}

\subsection{Walecka model}
\label{sec:Walecka}

The Walecka mean-field model of nuclear matter is the first and the simplest model from the quantum hadrodynamics framework. 
This model is based on the Lorentz invariant Lagrangian density, where  couplings of the nucleon field to the scalar $\sigma$ meson and to the vector $\omega$ meson fields are included,
and where the scalar field self interactions are neglected. 
The scalar and vector mesons act as
exchange particles, which mediate, respectively, the attractive and repulsive interactions between nucleons.
The mean field approximation prescribes that the meson fields and the nuclear current are replaced by their mean values, which are assumed spatially and time independent.
This enables to derive the pressure function, which plays the role of the grand canonical ensemble (GCE) thermodynamical potential, in the following form \cite{Walecka:1974qa}:
\eq{\label{p-wal}
p(T,\mu)~=~p^{\rm id}(T,\mu^*;m^*)~+~\frac{c^2_{\rm v}}{2}~n(T,\mu)~-~\frac{\left( m-m^*\right)^2}{2c^2_{\rm s}}~,
}
where $T$ and $\mu$ are, respectively, the system's temperature and chemical potential,
$n(T,\mu)$ is the nucleon density, 
$\mu^*$ and $m^*$ are, respectively, the effective nucleon chemical potential and mass.
$p^{\rm id}$ is the ideal Fermi pressure of a non-interacting nucleon gas and it is given by \Eq{p-id}.
The last two terms 
in Eq.~(\ref{p-wal}) 
correspond to the mean field contributions to the pressure
of the repulsive and attractive interactions between nucleons.
The model parameters $c^2_v$ and $c^2_s$ define the strength of the repulsive and attractive interactions, respectively.

The newly introduced quantities $\mu^*$ and $m^*$ are determined from the extremum condition
of the thermodynamical potential (i.e., maximum of the pressure):
\eq{
\label{mu-star-wal-therm}
& \left[ \frac{\partial p(T,\mu)}{\partial \mu^*}\right]_{T,m^* }\ =\ 0~,~~~~~~~~~
 \left[ \frac{\partial p(T,\mu)}{\partial m^*}\right]_{T,\mu }\ =\ 0~.
}
For the pressure~(\ref{p-wal}) this yields:
\eq{\label{mu-star-wal}
& \mu^*~=~\mu~-~c_{\rm v}^2~n(T,\mu)~,\\ \label{m-star-wal}
& \frac{m}{m^*}\, =\, 1\, +\, c_{\rm s}^2\frac{g_N}{2 \pi^2}
 \int_0^{\infty} k^2dk~\frac{ f_{\rm k}(T,\mu^*;m^*) }{\sqrt{ m^{*2}+{k}^2}} \,,
}
where $f_{\rm k}$ is the average occupation number, which corresponds to the momentum {\bf $k$}
and which is given by \Eq{quantum-f},
and $g_N \equiv 4$ is the spin-isospin nucleon degeneracy factor.
Equations (\ref{mu-star-wal}) and (\ref{m-star-wal}) are called gap equations.
In view of \Eq{mu-star-wal}, the Walecka model pressure
can be rewritten in the following symmetric form:
\eq{
p(T,\mu)~=~p^{\rm id}(T,\mu^*;m^*)~+~\frac{(\mu~-~\mu^*)^2}{2~c_{\rm v}^2}~-~\frac{(m~-~m^*)^2}{2~c_{\rm s}^2}~.
}
Using the standard thermodynamic relations in the GCE,
\eq{\label{therm-rel}
n(T,\mu) ~=~ \left[\der{p(T,\mu)}{\mu}\right]_T,~~~~~~~\varepsilon(T,\mu)~ =~T\left[\der{p(T,\mu)}{T}\right]_{\mu}~+~\mu~ n(T,\mu)~-~p(T,\mu)~,
}
one finds the nucleon number and energy density functions
of the Walecka model:
\eq{\label{n-wal}
& n(T,\mu)~=~n^{\rm id}(T,\mu^*;m^*)~,\\
\label{e-wal}
& \varepsilon(T,\mu)~=~\varepsilon^{\rm id}(T,\mu^*;m^*)~+~\frac{(\mu~-~\mu^*)^2}{2~c_{\rm v}^2}~+~\frac{(m~-~m^*)^2}{2~c_{\rm s}^2}~,
}
where the ideal gas particle density $n^{\rm id}$ and energy density $\varepsilon\id$ are given by Eqs.~(\ref{n-id}) and (\ref{e-id}), respectively.

\subsection{Quantum van der Waals model}
\label{sec:QvdW}

The classical van der Waals equation of state reads:
\eq{\label{classic-vdw-eos}
p(T,V,N)~=~\frac{T~N}{V~-~b~N}~-~a\left(\frac{N}{V}\right)^2~,
}
where $V$ is the system's volume, $N$ is the number of particles, $b~>~0$ is the excluded volume parameter characterizing the repulsive interactions, and $a~>~0$ is the parameter characterizing the strength of the attractive mean field.

The generalization of the vdW equation which includes the quantum statistical effects, and which is, therefore, suitable for the
description of nuclear matter, was
proposed in Ref.~\cite{Vovchenko:2015vxa}. 
The corresponding model will be referred to as the QvdW model. 
The GCE pressure and particle
density of the QvdW model are determined from the following system of two equations:
\eq{
p(T,\mu) & = p^{\rm id} (T, \mu^*) - a\,n^2(T,\mu)~,
\label{p-vdw}\\
n(T,\mu) & = \frac{n^{\rm id}(T,\mu^*)}{1 + b \, n^{\rm id}(T,\mu^*)}~,
\label{n-vdw}
}
where the effective chemical potential $\mu^*$ is given by
\begin{equation}
\label{mu-vdw}
\mu^* = \mu~ - ~b \, p(T,\mu) - a\,b\,n^2(T,\mu) + 2 \, a \, n(T,\mu)~.
\end{equation}
Here $p^{\rm id}$ and $n^{\rm id}$ are, respectively, the quantum
ideal gas pressure~(\ref{p-id}) and particle density~(\ref{n-id}).

The energy density in the QvdW model 
is calculated using the standard 
thermodynamical relations~(\ref{therm-rel}):
\eq{\label{e-vdw}
\varepsilon(T,\mu)~=~
\frac{\varepsilon^{\rm id}(T,\mu^*)}{1~+~b\,n^{\rm id}(T,\mu^*)}~-~ a\, n^2~
=~\left[\frac{\varepsilon^{\rm id}(T,\mu^*)}{n^{\rm id}(T,\mu^*)}
~-~a\,n\right]\,n~.
}

\subsection{Hybrid model}
\label{sec:Hybrid}

Let us additionally consider a \emph{Hybrid~model}, where 
the attractive interactions are mediated by the scalar $\sigma$ meson exchange, as in the Walecka model, and where the repulsive interactions are implemented by the excluded volume correction, as in the QvdW model. 
The equations for pressure and particle density within this model read
\eq{\label{p-hybrid}
p(T,\mu)\, & =\,  p^{\rm id}(T,\mu^*;m^*)~-~\frac{\left( m-m^*\right)^2}{2c^2_{\rm s}}~, \\
\label{n-hybrid}
n(T,\mu)\, &= \, \frac{n^{\rm id}(T,\mu^*;m^*)}{1~+~b~n^{\rm id}(T,\mu^*;m^*)}~,
}
where the effective chemical potential $\mu^*$ and the effective nucleons mass $m^*$ are determined by the following gap equations
\eq{\label{mu-hybrid}
& \mu^*~=~\mu~-~b~p^{\rm id}(T,\mu^*;m^*)~,\\
& \frac{m}{m^*}\, =\, 1\, +\, \frac{2 c_{\rm s}^2}{\pi^2}~ \big[1 + b\, n^{\rm id}(T,\mu^*;m^*)\big]^{-1}\,
\int_0^{\infty} k^2dk~\frac{ f(k,\mu^*;m^*) }{\sqrt{ m^{*2}+{k}^2}} \,,
\label{m-star-hybrid-new}
}
Finally, the energy density is obtained from the standard thermodynamic relations:
\eq{\label{e-hybrid}
\varepsilon~=~\frac{\varepsilon^{\rm id}(T,\mu^*;m^*)}{1~+~b~n^{\rm id}(T,\mu^*;m^*)}~+~\frac{\left( m-m^*\right)^2}{2c^2_{\rm s}}~.
}

\section{Nuclear matter}
\label{sec:nm}

In this section, the QvdW and Walecka models are used
to describe the properties of symmetric nuclear matter.
Our consideration will be restricted to small temperatures, $T\le 30$~MeV,
thus, a pion production is neglected.
In the present work, we also neglect a possible formation of nucleon clusters
(i.e., ordinary nuclei) and baryonic resonances (like $N^*$ and $\Delta$) which
may be important at low and high baryonic density, respectively.
Within these approximations, the number of nucleons $N$ becomes a conserved quantity,
and it plays the role of an independent variable in the canonical ensemble formulation.
In the GCE, the  chemical potential $\mu$ regulates the nucleon number density.
We fix the attractive and repulsive parameters for each model
in order to reproduce the properties of infinite
nuclear matter in its ground state at $T=0$ (see, e.g., Ref.~\cite{norm-nm}):
\eq{\label{propertiest0}
 p~=~0~,~~~~~~~~~~~~ \varepsilon / n ~\cong~ m + E_B ~\cong~ 922~{\rm MeV}~,~~~~~~~~~~~
 n~=~n_0 ~\cong ~ 0.16~{\rm fm}^{-3}~.
}
Here, $E_B \cong - 16$~MeV is the binding energy per nucleon.

We start with the QvdW model. The pressure $p$, the nucleon number
density $n$ and the energy density $\varepsilon$ are given by Eqs.~(\ref{p-vdw}~-~\ref{n-vdw}) and (\ref{e-vdw}),
respectively. 
One has the system of three equations (\ref{propertiest0})
for three unknown quantities, namely effective chemical potential at the ground state (GS) point $\mu^*_{\rm GS}$, $a$, and $b$. The solution to this system
of equations reads \cite{Vovchenko:2015pya}:
\eq{
{\rm QvdW:} \qquad
\mu^*_{\rm GS} ~ \cong  998~{\rm MeV}~,~~~~~~~~~b~ \cong  3.42 ~{\rm fm^3}~,~~~~~~~~~a ~ \cong ~ 329 ~{\rm MeV~fm^3}~.
}

For the Walecka model one has to substitute the corresponding expressions
for the thermodynamic quantities, namely  (\ref{p-wal}), (\ref{n-wal}), and (\ref{e-wal})
into Eq.~(\ref{propertiest0}). 
An additional expression (\ref{m-star-wal}) is also used to determine the effective mass at the GS point $m_{\rm GS}^*$.
At zero temperature, this expression reads
\begin{eqnarray}
\frac{m}{m_{\rm GS}^*}~=~
1\, +\, \frac{c_{\rm s}^2\, m_{\rm GS}^{*2}}
{\dis \pi^2}\,
\left[y \sqrt{y^2-1} -  \ln{\left(y + \sqrt{y^2-1}\right)}\right]~,
\label{eq:mstar-new55}
\end{eqnarray}
where $y \equiv \mu_{\rm GS}^*/m_{\rm GS}^*$.
Thus, one obtains the system of four equations for four unknowns, $\mu^*_{\rm GS}$, $m^*_{\rm GS}$, $c_{\rm v}^2$, and $c_{\rm s}^2$.
The solution to this system of equations yields\footnote{Note, that
our values of
parameters $c^2_s$ and $c^2_v$
are slightly different from those in Ref.~{\cite{Silva:2008zza}}. This is because
of a different ground state nuclear density $n_0\cong 0.15~{\rm fm}^{-3}$ used in {\cite{Silva:2008zza}}.}
\eq{
{\rm Walecka:} \quad
\mu^*_{\rm GS} ~ \cong 573~{\rm MeV}~,~~~~
m^*_{\rm GS} ~ \cong 510 ~{\rm MeV}~,~~~~
c_{\rm v}^2 ~ \cong ~ 11.0 ~{\rm fm^2}~,~~~~
c_{\rm s}^2 ~\cong ~ 14.6 ~{\rm fm^2}~.
}

In the Hybrid model one has to use expressions~\eqref{p-hybrid},\eqref{n-hybrid}, and \eqref{e-hybrid} when solving Eq.~(\ref{propertiest0}). One obtains:
\eq{
{\rm Hybrid:} \quad
\mu^*_{\rm GS} ~ \cong 897~{\rm MeV}~,~~~~
m^*_{\rm GS} ~ \cong 834 ~{\rm MeV}~,~~~~
b ~ \cong ~ 3.12 ~{\rm fm^3}~,~~~~
c_{\rm s}^2 ~\cong ~ 3.46 ~{\rm fm^2}~.
}

\begin{figure}[t]
\includegraphics[width=0.49\textwidth]{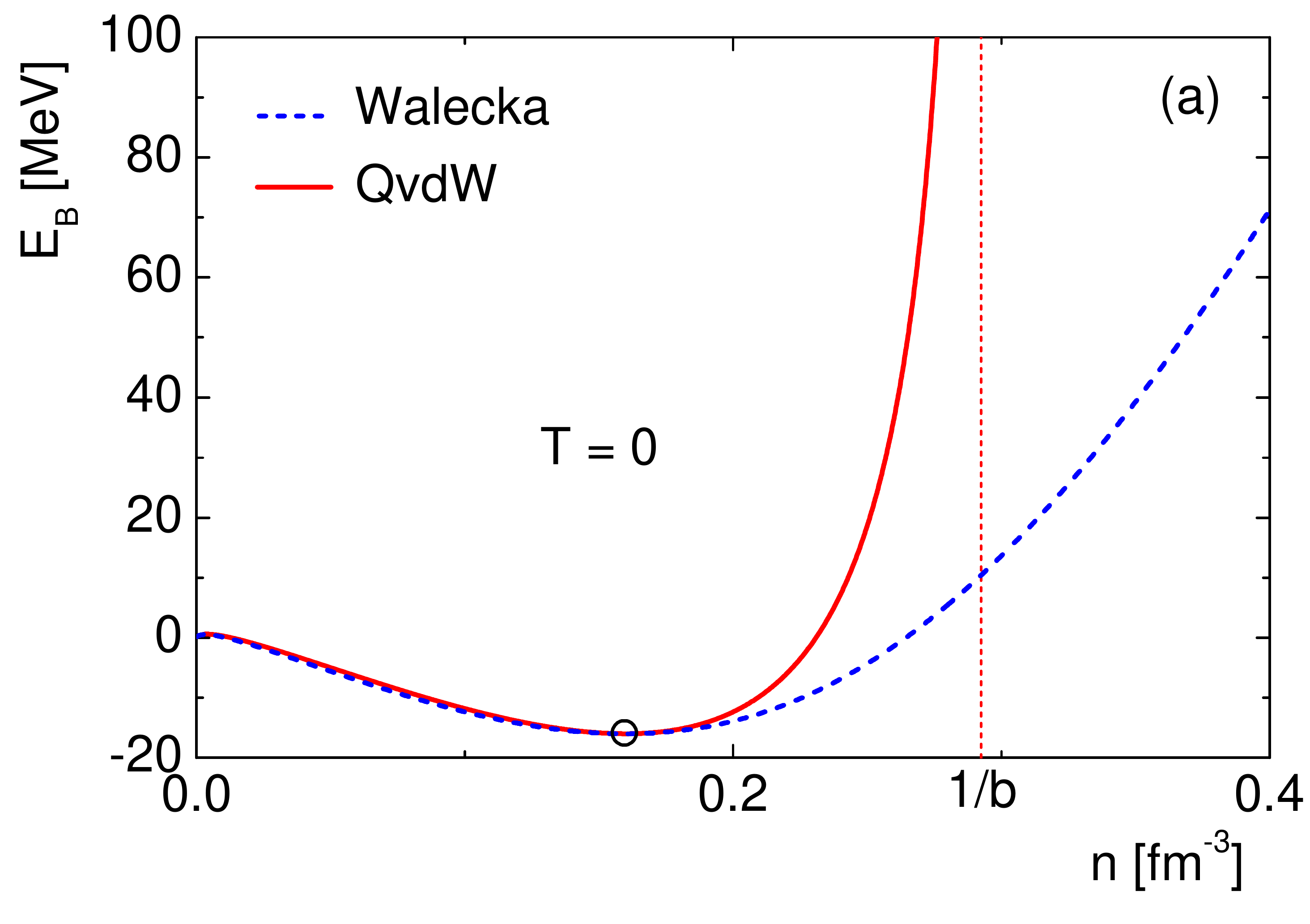}
\includegraphics[width=0.49\textwidth]{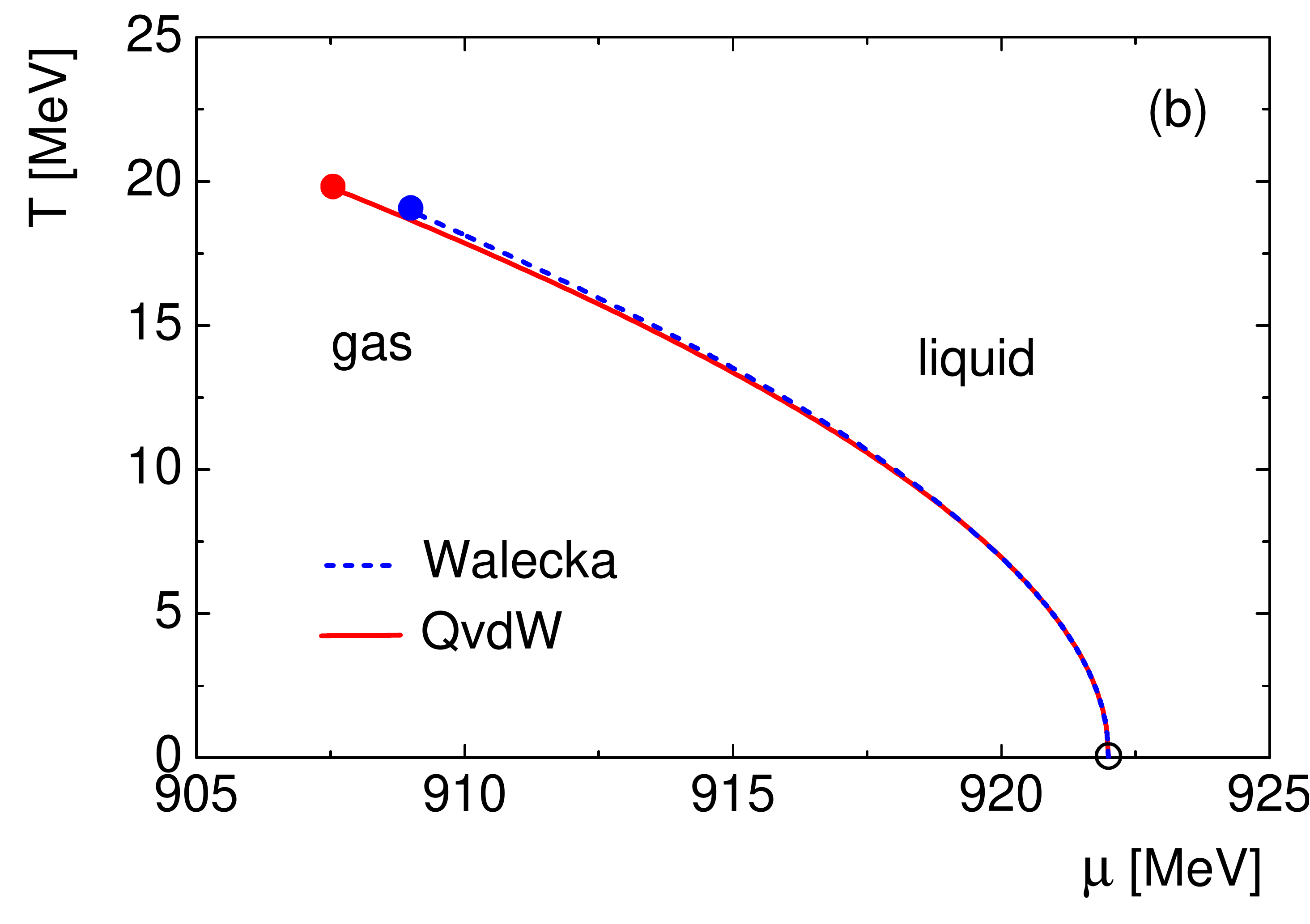}
\caption{$(a)$: Dependence of the binding energy per nucleon at $T=0$ in the Walecka~(dashed blue line) and the QvdW~(solid red line) models on the nucleon density.
The open circle depicts the nuclear ground state at $n = n_0 \cong 0.16~ {\rm fm}^{-3}$. 
The limiting density $n_{\rm lim} = 1/b$ in the QvdW model is shown by the vertical line.
$(b)$: Lines of the first-order liquid-gas phase transition in the $(\mu,T)$-plane for the Walecka~(dashed blue line) and the QvdW~(solid red line) models.
Solid circles at the end of each curve depict the CP in the corresponding model, while the the open circle depicts
the nuclear ground state.
}
\label{fig1}
\end{figure}

The dependence of the energy per nucleon on nucleon density at $T=0$ is shown in Fig.~\ref{fig1} $(a)$ for the Walecka and QvdW models.
At the ground state density $n = n_0 \cong 0.16~ {\rm fm}^{-3}$ both models yield $E_B = \varepsilon/n_0 - m = -16$~MeV by construction.
This point is depicted by an open circle. The results for $E_B$ at large densities differ considerably between the two models.
One sees that $E_B\rightarrow \infty$ at $n \rightarrow 1/b$ in the QvdW model. 
This is caused by the excluded volume repulsion, which leads to the upper  limit $n_{\rm lim}=1/b$ for the nucleon density. 
The same upper limit is present in the Hybrid model as
the repulsive interactions are also modeled with the vdW excluded volume correction.
In contrast, the nucleon density is not limited from above in the Walecka model. 
The applicability of the vdW excluded volume modeling is questionable at high densities in the vicinity of $n_{\rm lim}=1/b$.
We note that the range of the applicability can be extended to higher baryon densities by considering modifications of the vdW repulsive term.
One popular modification is the Carnahan-Starling model~\cite{CarnahanStarling}, which has a larger upper density limit of $n_{\rm lim}^{_{\rm CS}}=4/b$, and which has recently been successfully used in the hadronic physics applications~\cite{Satarov:2014voa,Anchishkin:2014hfa,Vovchenko:2017cbu}. 
The lines of the first-order liquid-gas phase transition in both models are shown in the $(\mu,T)$-plane in Fig.~\ref{fig1} (b). The end points of these lines correspond to the CP.
The two lines coincide at a point at $T=0$. This point corresponds to the nuclear ground state and it is depicted by the open circle.

At $T<T_c$ there is a first-order phase transition between the gas and liquid phases.
In order to find the gas and liquid densities $n_g$ and $n_l$ in the mixed phase at these temperatures, one applies the Maxwell rule of equal areas to the isotherms $p=p(v\equiv 1/n, T= {\rm const})$.
At $T = T_c$ these two densities coincide with the critical density $n_c$, which gives the location of the CP.
The locations of the CP  together with values for the incompressibility, $K_0 = 9 (dp/dn)_{T=0}$, within the Walecka, Hybrid and QvdW models
are presented in Table~\ref{tab1}. The experimental estimates~\cite{Elliott:2013pna} for these values are also shown in Table~\ref{tab1}.
All three considered models overshoot significantly the values of nuclear incompressibility $K_0$.
In that regard we note that the Walecka model
is the simplest model from the class of RMF models,
while the QvdW model is the simplest model from the class of real gas models.
Both models are known to overshoot empirical values of $K_0$. 
More elaborate models, such as the RMF Boguta-Bodmer model~\cite{Boguta:1977xi}, or the real gas Clausius model~\cite{Vovchenko:2017cbu}, can considered to address this issue.
We do not expect such modifications to change qualitative features of the critical behavior in nuclear matter, which are studied in this work.

The values of $T_c$, $n_c$, $p_c$ and $K_0$
within the Hybrid model are located between the corresponding values within the Walecka and QvdW models. 
Since all the Hybrid model results are intermediate between the Walecka and QvdW ones, they are not shown in the figures.
\begin{figure}[t]
\includegraphics[width=0.49\textwidth]{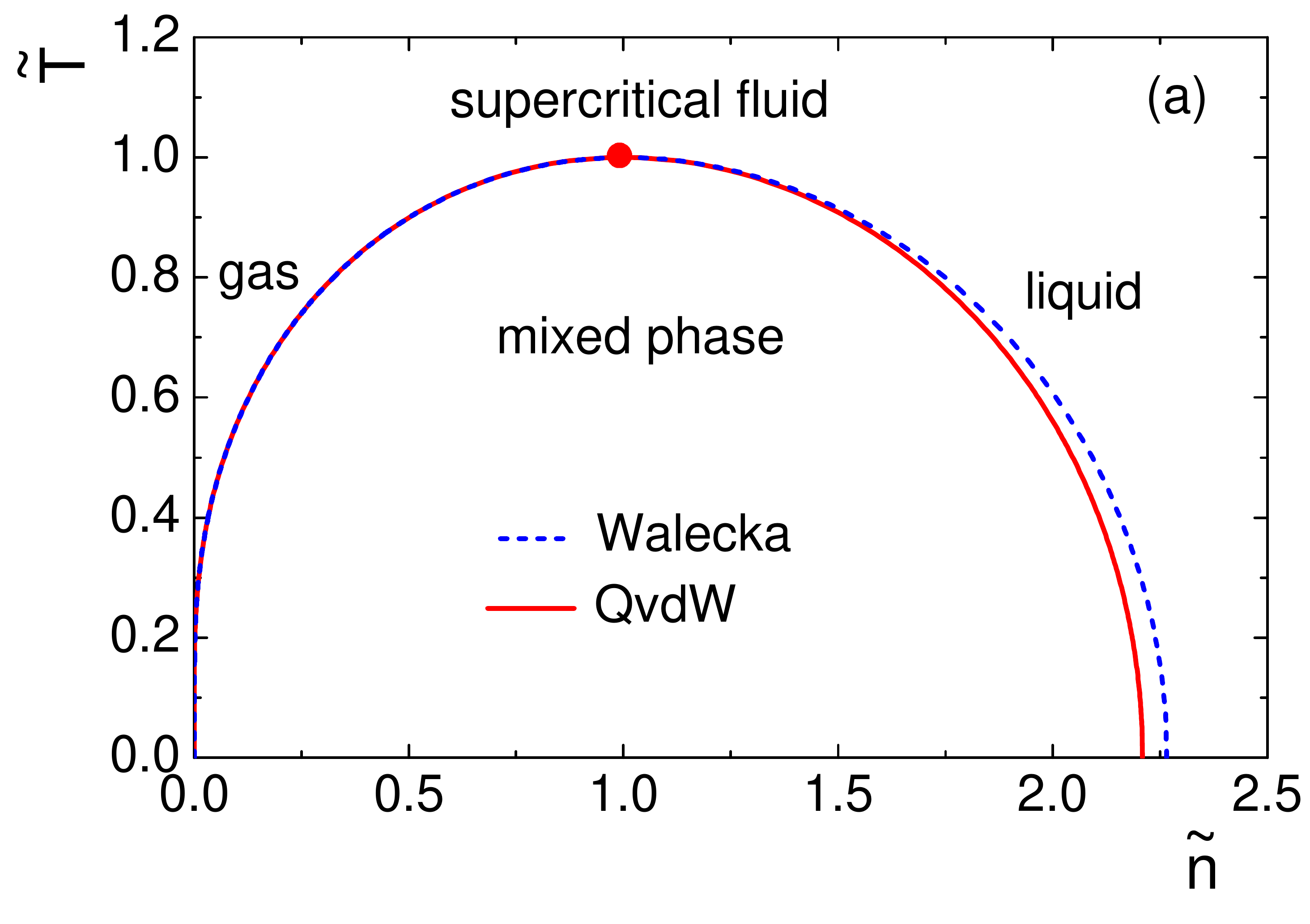}
\includegraphics[width=0.49\textwidth]{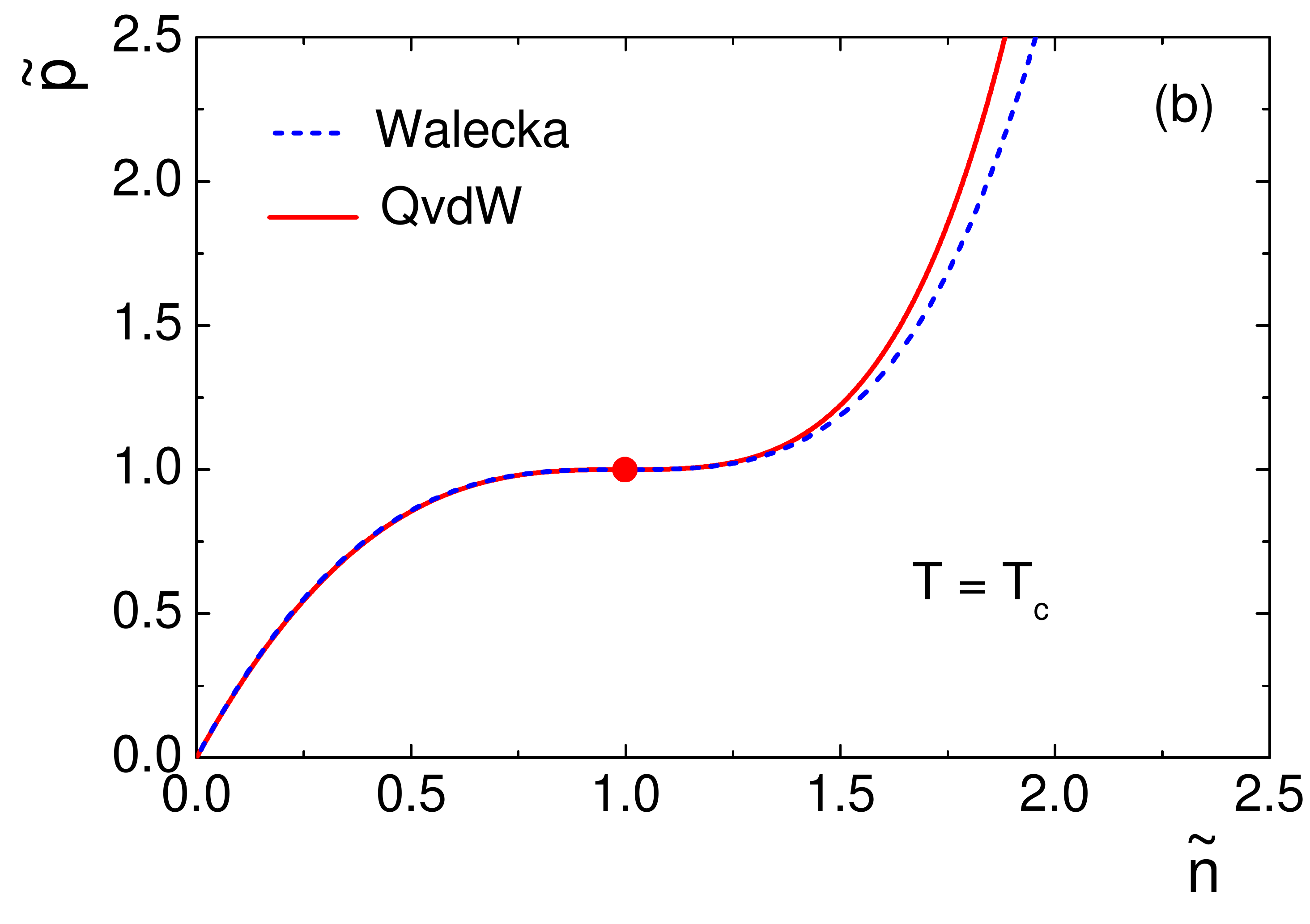}
\caption{\label{coex_curves}
The scaled coexistence curves $(a)$ and the scaled critical isotherms $(b)$ calculated for the Walecka~(dashed blue lines) and the QvdW~(solid red lines) models. 
The solid circles correspond to the CP.
}
\end{figure}

\tab{
{| l | c | c | c | c | }
\hline
\hline
 &                          ~Walecka~  &~~Hybrid~~~&~~QvdW~~ &~Experiment \cite{Elliott:2013pna}~ \\
\hline
 ~$T_c~[\rm{MeV}]$               & 18.9  & 19.2 & 19.7 & $17.9~\pm~0.4$\\
 ~$n_c~[\rm{fm}^{-3}]$           &0.070  & 0.071 &0.072 & $0.06~\pm~0.01$\\
 ~$p_c~[\rm{MeV}\,\rm{fm}^{-3}]$~  & 0.48  & 0.50 &0.52 & $0.31~\pm~0.07$ \\
~$K_0$   [\rm{MeV}]                 & 553   & 674  & 763  & 250~-~315 \\
  \hline
   \hline
}{\label{tab1}
The location of the CP and the value of the ground state incompressibility $K_0$ within the Walecka, Hybrid and QvdW models, together with their experimental estimates.
}

Let us introduce the following reduced variables: $\widetilde{T} \equiv T/T_c$,  $\widetilde{n} \equiv n/n_c$, and $\widetilde{p} \equiv p/p_c$.
The classical vdW equation of state (\ref{classic-vdw-eos}) becomes
independent of the 
interaction parameters $a$
and $b$ when expressed in these reduced variables.
This independence on the interaction parameters is known as the $law~of~corresponding~states$. 
Although vdW equation do not describe real gases and liquids such a law is known to hold true for many real gases and liquids with rather good accuracy (see, e.g., Ref.~\cite{balescu}).
The coexistence curves and critical isotherms in the reduced coordinates for both models are presented in Figs.~\ref{coex_curves} $(a)$ and
$(b)$, respectively. One sees
only a tiny 
difference between the two models, mostly in the high density region.

\section{Critical exponents}
\label{sec:cr}

$Critical~exponents$ 
characterize the critical behavior of the thermodynamic quantities in the vicinity of the CP. 
This is done in terms of the power law expressions (see, e.g., Ref.~\cite{balescu}).
The so-called thermodynamic critical exponents are $\alpha$, $\beta$, $\gamma$, and $\delta$.
Their definitions are presented in the first two columns of Table~\ref{crit-exp-tab}.

\tab{
{| c | c | c || c | c | }
\hline
\hline
~~~~~~ & 1 & 2 & 3 & 4  \\
\hline
exponent & scaling & path & mean-field theory & empirical  \\
\hline
$\alpha$    & $c_{\rm v}~=~A~\tau^{-\alpha}$ &  ~$\tau~\rightarrow~0^+~,~\widetilde{n}~=~1$~ & 0
& 0.11\\
$\beta$     & $\ddfrac{\wt{n}_{\rm l}~-~\wt{n}_{\rm g}}{2}=~B~(-\tau)^{\beta}$ &   ~$\tau~\rightarrow~0^-$~  & $\ddfrac{1}{2}$
& 0.33\\
$\gamma$     & $\kappa_T~=~p_c^{-1}~G~\tau^{-\gamma}$ &   ~$\tau~\rightarrow~0^+~,~\widetilde{n}~=~1$~ & 1
& 1.24\\
$\delta$     & $~\wt{p}~-~1~=~D~|\wt{n}~-~1|^{\delta}~{\rm sgn}(\wt{n}~-~1)~$ &   ~$n~\rightarrow~1^\pm$~,~$\tau~=~0~$~ & 3
& 4.79\\
 \hline
 \hline
}{\label{crit-exp-tab} The thermodynamic critical exponents. Column (1): The scaling relations.
Here $\tau = \wt{T}-1$, $c_{\rm v}$ is a specific heat capacity at constant density.
The order parameter $\widetilde{n}_{\rm l}-\widetilde{n}_{\rm g}$ is a width 
of the mixed phase region  at a given $\wt{T}$. Isothermal compressibility is $\kappa_T = n^{-1}[\partial n/\partial p]_T$.
(2): The thermodynamic path of approach of CP. (3): Values of the critical  exponents in the universality class of the mean-field theory. 
(4): 
Empirical values of the critical exponents for real gases~\cite{3dising}.}

\begin{figure}[!htb]
\includegraphics[width=0.49\textwidth]{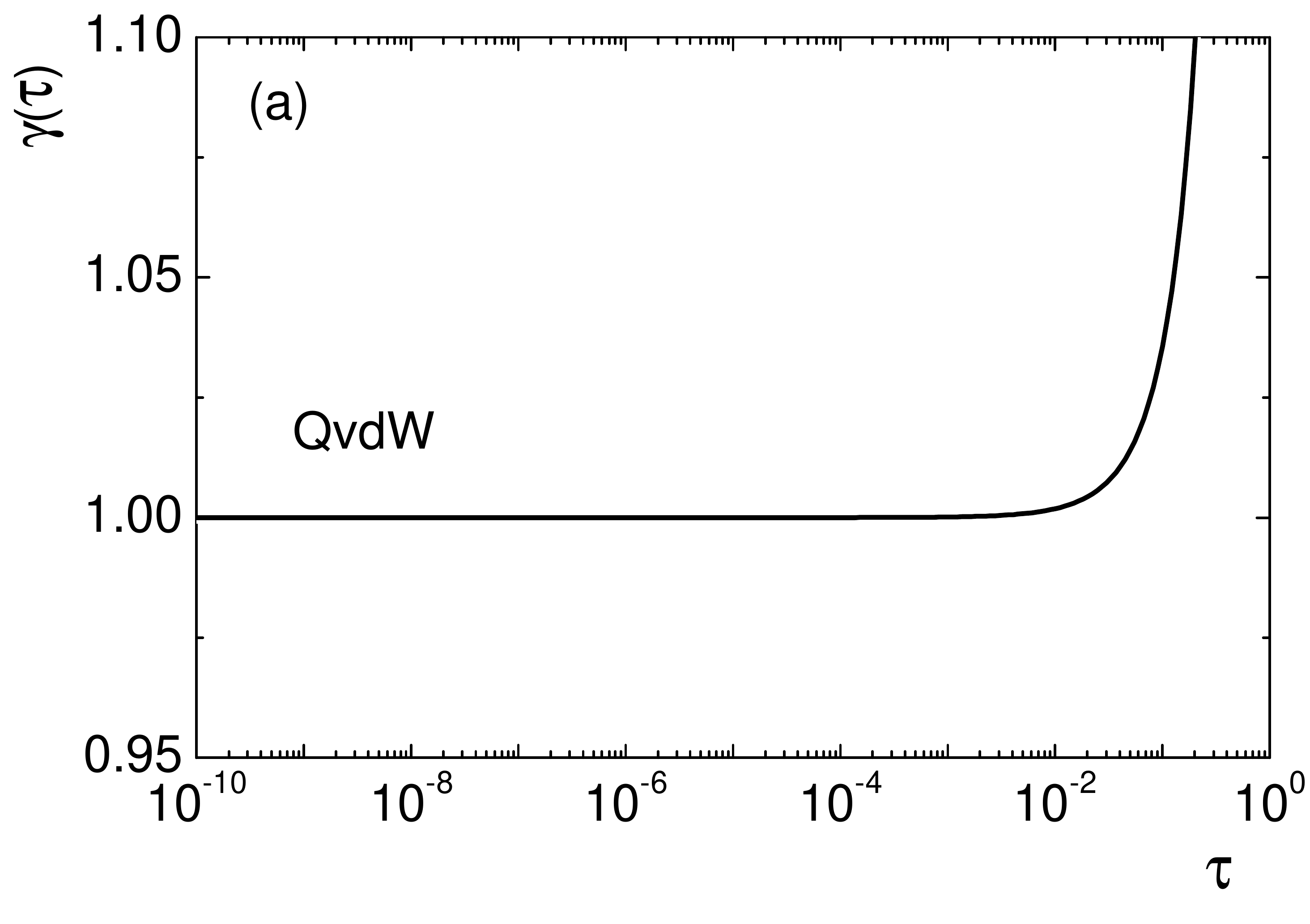}
\includegraphics[width=0.49\textwidth]{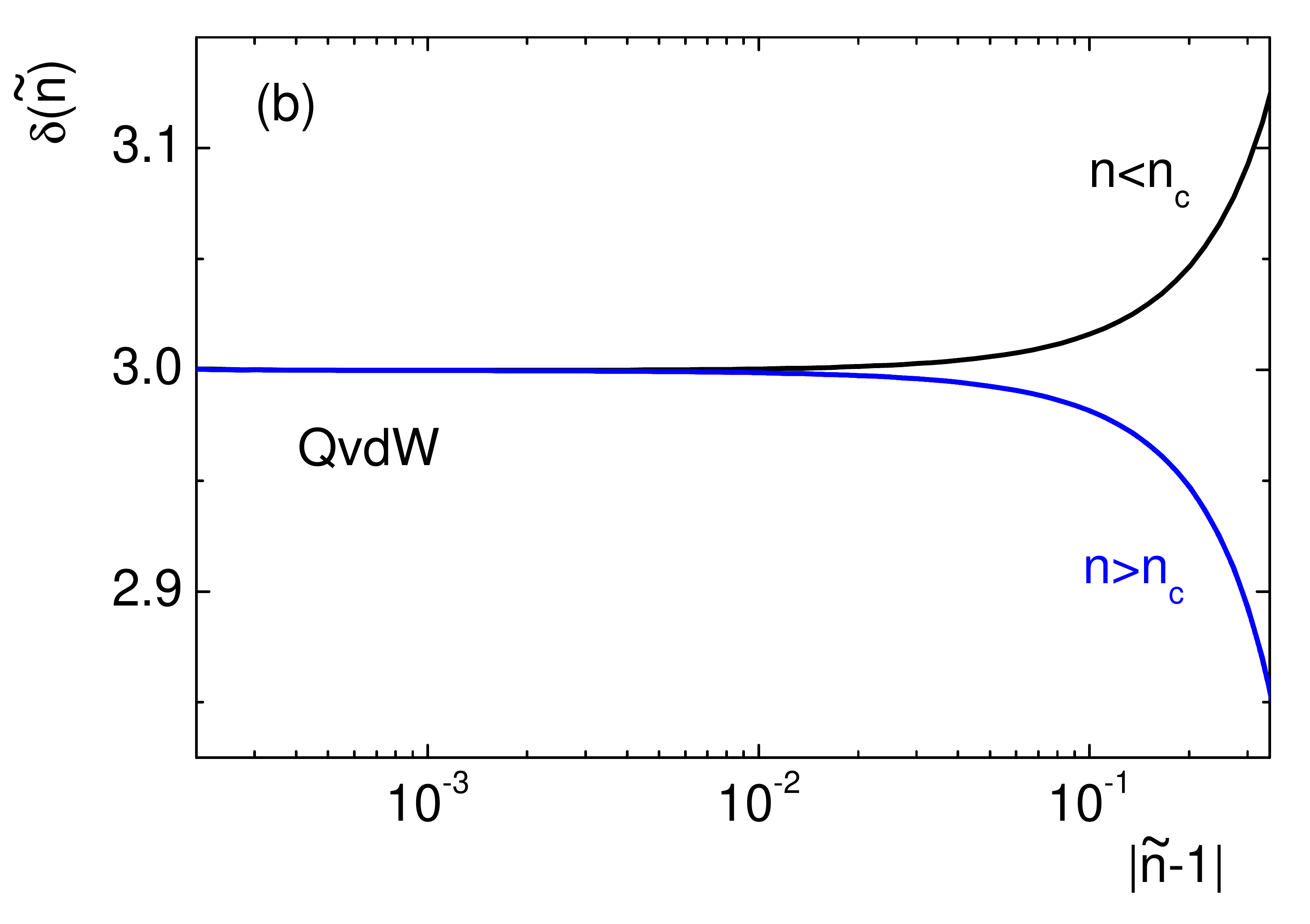}
\caption{\label{crit-exp}
The dependence of the $\gamma(\tau)$ $(a)$ and $\delta(\widetilde{n})$  $(b)$ exponents as functions of a proximity to the CP. 
}
\end{figure}

The critical behavior in the Walecka model had been studied before~(see, e.g., Ref.\cite{Silva:2008zza}). 
The obtained values for the critical exponents are $\alpha=0$, $\beta=1/2$, $\gamma=1$, and $\delta=3$. 
These values of exponents correspond to the universality class of the \emph{mean-field} theory. A variety of models belong to this class, including, e.g., the classical vdW model.
The critical exponents within the QvdW model 
are expected to
satisfy the \emph{scaling relations} 
\eq{\label{scaling}
\alpha~+~2~\beta~+~\gamma~=~2~,~~~~~~~~~~~{\rm and}~~~~~~~~~~~~\gamma~=~\beta(\delta~-~1)~,
}
which are called the Rushbrook and Widon scaling relation, respectively (see, e.g., Ref.~\cite{balescu}). Thus, it is sufficient to determine only a pair of critical exponents.

First, let us find the $\gamma$ critical exponent.
It can be shown that the isothermal compressibility $\kappa_T$ is proportional to the  scaled variance of the particle number distribution, i.e.
$T~n~\kappa_T = \omega[N]$.
Thus, at $\tau = \tilde{T} - 1 ~\rightarrow~0^{+}$ and $\wt{n}~=~1$ one has
\eq{\label{w-gamma}
\omega[N]~=~
\frac{T_c~n_c}{p_c}~G~ \tau^{-\gamma}~,
}
where $G$ is the critical amplitude.
The scaled variance 
in the QvdW model reads~\cite{Vovchenko:2015pya}
\eq{\label{w-vdw}
\omega[N]~\equiv ~\frac{\mean{N^2}-\mean{N}^2}{\mean{N}}~=~\frac{T}{n}\left(\der{n}{\mu}\right)_T~=~
\omega^{\rm id}(T,\mu^*)\left[\frac{1}{1~-~bn^2}~-~\frac{2an}{T}~\omega^{\rm id}(T,\mu^*)\right]^{-1}~,
}
where $\omega^{\rm id}(T,\mu^*)$ is the scaled variance for the corresponding ideal gas~(\ref{w-id}).
The factor in the square brackets in the r.h.s. of Eq.~\eqref{w-vdw} goes to zero at $T \rightarrow ~T_c$. As the result, $\omega[N]$ diverges~\cite{Vovchenko:2015pya} at the CP. 
Let us perform the Taylor series expansion of this factor in the vicinity of the critical temperature $T_c$ for $T > T_c$ at the critical density $n=n_c$. 
Neglecting the terms proportional to the second or higher power of $\tau$, one obtains
\eq{
\left[\frac{1}{1~-~bn^2}~-~\frac{2an}{T}~\omega^{\rm id}\right]
 ~\cong ~
-~\frac{2~a~n_c}{T_c}\left[T_c\left(\dfrac{d\omega^{\rm id}}{dT}\right)-~\omega^{\rm id}\right]_{n_c,T_c}~\tau~.\label{factor}
}
where the $(d\omega^{\rm id}/dT)$ derivative is taken at a constant particle number density $n = n_c$.
Substituting this result into \Eq{w-vdw} 
one obtains the following expression
\eq{\label{w-vdw-crit}
\omega[N]~=~-\frac{T_c}{2~a~n_c}\left[\frac{\omega^{\rm id}}{T_c(d\omega^{\rm id}/dT)-\omega^{\rm id}}\right]_{n_c,T_c}\tau^{-1}~.
}
Comparing \Eq{w-vdw-crit} with \Eq{w-gamma} one gets the value of the critical exponent of $\gamma=1$ and the critical amplitude 
\eq{\label{g}
G~=~-~\frac{p_c}{2~ a~ n_c^2}\left[\frac{\omega^{\rm id}}{T_c(d\omega^{\rm id}/dT)-\omega^{\rm id}}\right]_{n_c,T_c}~\cong~0.257~.
}
The value of $\gamma$ is obtained analytically while the value of $G$ is calculated numerically. 
Note that in the Boltzmann approximation $\omega^{\rm id}\equiv 1$, and Eq.~(\ref{g}) reads
$G=p_c/(2an_c^2)$. Using the classical vdW expressions $p_c=a/27b^2$ and $n_c=1/3b$ for the critical parameters~\cite{balescu}, one finds that $G=1/6$ for the classical vdW model. 
Thus, the deviation of the critical amplitude $G$ in the QvdW model from its classical vdW value is the effect of Fermi statistics.
On the other hand, the $\gamma = 1$ value is the same in both, the classical and the quantum vdW models.

The definition of the $\gamma$ critical exponent can be also written in the following form,
\eq{\label{gamma-def}
\gamma ~= ~\lim_{\tau \rightarrow + 0} \left[\gamma(\tau)\right]_{\widetilde{n} = 1}~,~~~{\rm where}~~~\gamma(\tau)~
=~-\frac{{\rm ln}~\omega[N]~+~{\rm ln}\left[\dfrac{T}{2an_c}~G\right]}{{\rm ln}~\tau}~.
}
This particular form is quite general, and it is useful for the numerical determination of $\gamma$.
To illustrate this, we depict the $\gamma(\tau)$  function
in Fig. \ref{crit-exp} $(a)$ for the QvdW model of nuclear matter. 
One sees that, for $\tau \lesssim 0.01$,  $\gamma(\tau)$ is already not distinguishable from unity.

We consider only numerical computation of the $\delta$ critical exponent in the QvdW model of nuclear matter.
In analogy to Eq.~\eqref{gamma-def},
the $\delta$ exponent~(see Table \ref{crit-exp-tab}), can be rewritten as
\eq{\label{delta-def}
\delta ~= ~\lim_{\widetilde{n} \rightarrow 1} \left[\delta(\widetilde{n})\right]_{\widetilde{T} = 1}~,~~~
{\rm where}~~~\delta(\widetilde{n})~=~\frac{{\rm ln}|\wt{p}~-~1|~+~{\rm ln}D}{{\rm ln}|\wt{n}~-~1|}~,
}
for both $\tilde{n} < 1$ and $\tilde{n} > 1$.

The dependence $\delta(\widetilde{n})$ at the critical isotherm $T=T_c$ is presented in Fig.~\ref{crit-exp}~$(b)$ 
for $n<n_c$ (upper line) and $n>n_c$ (lower line).
The numerical calculation yields $\delta=3$ and 
\eq{
D ~= ~\lim_{\widetilde{n} \rightarrow 1} 
\left[\frac{|\wt{p}~-~1|}{|\wt{n}~-~1|^3} ~{\rm sgn}(\wt{n}~-~1)\right]_{\widetilde{T} = 1}~\cong~1.4~.
}

The $\delta = 3$ value coincides with the one predicted by the classical vdW model. 
The critical amplitude $D$ differs from the classical vdW value of $D=3/2$ due to the effects of Fermi statistics.
It is seen in Fig. \ref{crit-exp} $(b)$ that 
$\delta(\wt{n})$ is not distinguishable from 3 at $|\wt{n}~-~1| \lesssim 0.01$, for both $n>n_c$ and $n<n_c$.

The scaling relations~(\ref{scaling}) 
yield the values of the remaining critical exponents for the QvdW model: $\alpha = 0$, $\beta = 1/2$.
All values of the critical exponents, obtained for the QvdW model,
coincide with the values given by the mean-field universality class.
Note, however, that the values of corresponding critical amplitudes are not universal and remain different in different models.

\section{Summary}
\label{sec:summary}

An infinite system of interacting nucleons
(nuclear matter) has been considered in the framework of 
the Walecka model \cite{Walecka:1974qa} and of the quantum formulation of the van der Waals model \cite{Vovchenko:2015vxa}.
A third model discussed in the present paper 
is the Hybrid model, where 
the attractive interactions are mediated by the scalar $\sigma$ meson exchange, as in the Walecka model, and where the repulsive interactions are implemented by the excluded volume correction, as in the quantum van der Waals model.

Each model contains two parameters which characterize the repulsive and attractive interactions between nucleons.
These parameters are determined in each model in order to reproduce the properties of the ground state of nuclear matter.
All considered models predict the existence of a liquid-gas
phase transition and a critical point.
The scaled coexistence curves and the scaled critical isotherms are very similar in the Walecka and quantum van der Waals models. Both models have identical critical exponents which coincide with those of the classical van der Waals model. Therefore, both models
fall into the mean-field universality class.  
The Hybrid model leads to the results which lie ``in-between'' the Walecka and quantum van der Waals model predictions.
It is notable that the critical amplitudes, which characterize the critical behavior, are different in the quantum and in the classical van der Waals models.

Despite strong similarity of the considered models in the vicinity of the critical point, there are important differences in the behavior of their thermodynamic functions in other part of the phase diagram. 
A qualitative difference takes place at large values
of the nucleon number density $n$: both the quantum van der Waals and hybrid models have an upper density limit of $n< 1/b$, whereas there is no such restriction in the Walecka model.

\section*{Acknowledgments}
We are thankful to Leonid Satarov and Horst Stoecker for fruitful discussions. This work is supported by  the Program of Fundamental Research of the Department of Physics and
Astronomy of National Academy of Sciences of~Ukraine.
The work of M. I. G. is also supported by the
Goal-Oriented Program of  the National Academy of Sciences of Ukraine, by the
European Organization for Nuclear Research (CERN), Grant CO-1-3-2016.

\appendix
\section{Ideal Fermi Gas}
The pressure function 
of the ideal Fermi gas
reads
\eq{
p^{\rm id}(T,\mu;m)~ &=~
 \frac{ g}{6\pi^2} \int_0^{\infty} k^2\,dk
\frac{ k^2}{\sqrt{m^{2} + k^2}}\, f_{\rm k}(T,\mu; m)  \,,
\label{p-id}
}
where 
\eq{
 f_{\rm k}(T,\mu; m)\, = \, 
\left[\exp{ \left( \frac{\sqrt{ m^{2}+k^2} - \mu}{T}\right)} + 1\right]^{-1}\,,
\label{quantum-f}
}
$g$ is the degeneracy factor and
$m$ is the particle's mass~(for nucleons one has $g = 4$ and $m = 938$~MeV).
The  particle number density $n^{\rm id}$
and energy density $\varepsilon^{\rm id}$ are obtained
using the standard thermodynamical relations
\eq{\label{n-id}
& n^{\rm id}(T,\mu;m)~=~ \left[\der{p^{\rm id}}{\mu}\right]_T~=~
 \frac{ g}{2\pi^2} \int_0^{\infty} k^2\,dk
\, f_{\rm k}(T,\mu; m)  \,,\\\label{e-id}
& \varepsilon^{\rm id}(T,\mu;m)=~T\left[\der{p^{\rm id}}{T}\right]_{\mu}~+~\mu~ n^{\rm id}~-~p^{\rm id}~=~
 \frac{ g}{2\pi^2} \int_0^{\infty} k^2\,dk
\sqrt{m^{2} + k^2}\, f_{\rm k}(T,\mu; m)  \,.
}

At zero temperature the ideal gas quantities
can be written as
\begin{eqnarray}
\label{eq:n-id-t0}
n^{\rm id}(T=0,\mu; m) &=& \frac{g}{2\pi^2} \int\limits_0^{\sqrt{\mu^{2}-m^{2}}}
dk\, k^2\,
=\, \frac{g}{6\pi^2} \, (\mu^{2}-m^{2})^{3/2}\,
=\, \frac{g\,m^{3}}{6\pi^2} \, (y^2-1)^{3/2}\,,
\\
\label{eq:p-id-t0}
p^{\rm id}(T=0,\mu; m) &=& \frac{g}{6 \pi^2} \int\limits_0^{\sqrt{\mu^{2}-m^{2}}} dk\,
\frac{k^4}{\sqrt{k^2+m^{2}}}\,
=\, \frac{g\,m^{4}}{6 \pi^2} \int\limits_0^{\sqrt{y^2-1}} dx\,
\frac{x^4}{\sqrt{x^2+1}}\, =
\nonumber \\
&=& \frac{g\,m^{4}}{48 \pi^2}\, \left[y \sqrt{y^2-1} (-5 + 2y^2)
+ 3 \ln{\left(y + \sqrt{y^2-1}\right)}\right]\,,
\\
\label{eq:e-id-t0}
\varepsilon^{\rm id}(T=0, \mu; m) &=& \frac{g}{2 \pi^2}
\int\limits_0^{\sqrt{\mu^{2}-m^{2}}} dk \, k^2\, \sqrt{k^2+m^{2}}\,=
\frac{g\,m^{4}}{2\pi^2}
\int\limits_0^{\sqrt{y^2-1}} dx \, x^2\, \sqrt{x^2+1}\,=
\nonumber \\
&=& \frac{g\,m^{4}}{16\pi^2} \left[ y \sqrt{y^2-1} (2 y^2 - 1)
- \ln{\left(y + \sqrt{y^2-1}\right)} \right] \,,
\\
s^{\rm id} (T=0,\mu; m) &=& \lim_{T \to 0} \frac{\varepsilon^{\rm id}(m,T, \mu)
+ p^{\rm id}(m,T,\mu) - \mu \, n^{\rm id}(m,T, \mu)}{T}~ =~ 0~,
\label{eq:s-id-t0}
\end{eqnarray}
where $y \equiv \mu/m$.

The scaled variance of particle number distribution in the ideal Fermi gas reads
\eq{\label{w-id}
\omega^{\rm id}(T,\mu;m)~=~\frac{\mean{N^2}-\mean{N}^2}{\mean{N}}~=~\frac{T}{n^{\rm id}}\left(\der{n^{\rm id}}{\mu}\right)_T~=~1~-~\frac{g}{2 \pi^2 n^{\rm id}}\int_0^{\infty}dk k^2 f^2_{\rm k}(T,\mu; m)~,
}
where $n^{\rm id}$ is given by \Eq{n-id}. Note that in the Boltzmann approximation $\omega^{\rm id}\equiv 1$~.


\end{document}